%% file: PSCC.tex
\documentclass{IEEEtran4PSCC}[twocolumn]
\ifCLASSINFOpdf
   \usepackage[pdftex]{graphicx}
\else
   \usepackage[dvips]{graphicx}
\fi
\usepackage{amssymb}
\usepackage{mathtools}
\usepackage{subcaption}
\usepackage{svg}
\usepackage{float}
\usepackage{eurosym}
\usepackage{comment}
\usepackage[]{amsmath}
\usepackage[ruled, vlined, linesnumbered]{algorithm2e}
\newcommand{\nonl}{\renewcommand{\nl}{\let\nl\oldnl}}

\usepackage{xcolor}
\usepackage{cleveref}
\usepackage{booktabs}
\usepackage{array}
\usepackage{flushend}

\usepackage{multicol}
\DeclareMathOperator*{\argmax}{arg\,max}
\makeatletter
\let\old@ps@headings\ps@headings
\let\old@ps@IEEEtitlepagestyle\ps@IEEEtitlepagestyle
\def\psccfooter#1{%
    \def\ps@headings{%
        \old@ps@headings%
        \def\@oddfoot{\strut\hfill#1\hfill\strut}%
        \def\@evenfoot{\strut\hfill#1\hfill\strut}%
    }%
    \def\ps@IEEEtitlepagestyle{%
        \old@ps@IEEEtitlepagestyle%
        \def\@oddfoot{\strut\hfill#1\hfill\strut}%
        \def\@evenfoot{\strut\hfill#1\hfill\strut}%
    }%
    \ps@headings%
}
\makeatother

\psccfooter{%
        \parbox{\textwidth}{\hrulefill \\ \small{24th Power Systems Computation Conference} \hfill \begin{minipage}{0.2\textwidth}\centering \vspace*{4pt} \includegraphics[scale=0.06]{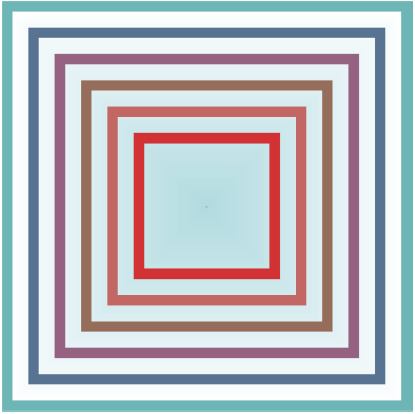}\\\small{PSCC 2026} \end{minipage} \hfill \small{Limassol, Cyprus --- June 8 -- June 12, 2026}}%
}

\begin{document}
%


\title{Explicit Reward Mechanisms for Local Flexibility in Renewable Energy Communities}

\author{
\IEEEauthorblockN{Thomas Stegen$^{1}$, Julien Allard$^{1}$, Noé Diffels$^{1,2}$, François Vallée$^{1}$, Mevludin Glavic$^{2}$,\\ Zacharie De Grève$^{1}$ and Bertrand Cornélusse$^{2}$\\}
\IEEEauthorblockA{$^{1}$Power Systems and Markets Research Group, University of Mons, Belgium \\ $^{2}$Montefiore Institute, University of Liège, Belgium} 
}

\include{macros}

\maketitle

\begin{abstract}
Incentivizing flexible consumption of end-users is key to maximizing the value of local exchanges within Renewable Energy Communities. If centralized coordination for flexible resources planning raises concerns regarding data privacy and fair benefits distribution, state-of-the-art approaches (\textit{e.g.}, bi-level, ADMM) often face computational complexity and convexity challenges, limiting the precision of embedded flexible models. 
This work proposes an iterative resolution procedure to solve the decentralized flexibility planning with a central operator as a coordinator within a community. The community operator asks for upward or downward flexibility depending on the global needs, while members can individually react with an offer for flexible capacity. This approach ensures individual optimality while converging towards a global optimum, as validated on a 20-member domestic case study for which the gap in terms of collective bill is not more than 3.5\% between the decentralized and centralized coordination schemes.
\end{abstract}


\begin{table}[H]
\caption{Variables of the model}
\centering
\begin{tabular}{l|c|c|c|p{.38\columnwidth}}
& Dim & Dom & Unit & Description \\\hline
\rule{0pt}{2.5ex}\exportCom & \UxT & \Rplus & kW & Community export power\\
\importCom & \UxT & \Rplus & kW & Community import power\\
\exportGrid & \UxT & \Rplus & kW & Retailer export power\\
\importGrid & \UxT & \Rplus & kW & Retailer import power\\
\PPV & \UxT & \Rplus & kW & PV production power\\
\charge & \UxT & \Rplus & kW & BSS charging power\\
\discharge & \UxT & \Rplus & kW & BSS discharging power\\
\Pinj & \UxT & \Rplus & kW & Node injection power\\
\flex & \UxT & \Rplus & kW & Flexible load power\\
\Pow{EV} & \UxT & \Rplus & kW & EV charging power\\
\Pow{WB} & \UxT & \Rplus & kW & WB power setpoint\\
\Pow{HP} & \UxT & \Rplus & kW & HP power setpoint\\
\SoC{EV} & \UxT & $\left[0,1\right]$ & - & EV state of charge\\
\Temp{WB} & \UxT & \Rplus & \degC & WB tank temperature\\
\Temp{HP} & \UxT & \Rplus & \degC & HP indoor temperature\\
\discomf{EV} & \UxT & \Rplus & \euro & Discomfort from EV flex\\
\discomf{WB} & \UxT & \Rplus & \euro & Discomfort from WB flex\\
\discomf{HP} & \UxT & \Rplus & \euro & Discomfort from HP flex\\
\end{tabular}
\label{tab:variables}
\end{table}

\begin{table}[htpb]
\caption{Inputs or parameters of the model}
\centering
\begin{tabular}{l|c|c|c|p{.37\columnwidth}}
& Dim & Dom & Unit & Description \\\hline
\rule{0pt}{2ex}\fix & \UxT & \Rplus & kW & Fixed power consumption\\
\PPVprof & \UxT & \Rplus & kW & Max PV power production\\
\Capa{PV} & $\mathcal{U}$ & \Rplus & kWp & PV nominal power\\
\SoC{init,BSS} & $\mathcal{U}$ & $\left[0,1\right]$ & - & Initial battery state of charge\\
\Capa{BSS} & $\mathcal{U}$ & \Rplus & kWh & Battery capacity\\
\eff{BSS} & $\mathcal{U}$ & $\left[0,1\right]$ & - & Battery efficiency\\
\price{i,ret} & $\mathcal{T}$ & \Rplus & \euro/kWh & Retailer import price\\
\price{e,ret} & $\mathcal{T}$ & \Rplus & \euro/kWh & Retailer export revenue\\
\OPFee & $\mathcal{T}$ & \Rplus & \euro/kWh& Community fee\\
\Pow{ref} & $\UxT$ & \Rplus & kW & Reference flexible profile\\
$E^{\text{Cap}}$ & $\mathcal{U}$ & \Rplus & kWh & Total flexible load\\
\hline
\multicolumn{5}{l}{\rule{0pt}{2ex}\textbf{Electric Vehicle}} \\\hline
\rule{0pt}{2.5ex}${a}^{\text{EV}}$ & \UxT & \Bool & - & EV plugged\\
${t}^{\text{arr,EV}} $& \UxT & \Bool & - & EV arriving\\
${t}^{\text{dep,EV}} $& \UxT & \Bool & - & EV leaving\\
\SoC{arr,EV} & \UxT & $\left[0,1\right]$ & - & EV state of charge at arrival\\
\SoC{ref,EV} & \UxT & $\left[0,1\right]$ & - & EV reference state of charge\\
\Pow{ref,EV} & \UxT & \Rplus & kW & Power from reference profile\\
$\bar{s}^{\text{EV}}$ & $\mathcal{U}$ & $\left[0,1\right]$ & - & Target state of charge\\
\SoC{init,EV} & $\mathcal{U}$ & $\left[0,1\right]$ & - & Initial state of charge\\
\eff{EV}& $\mathcal{U}$ & $\left[0,1\right]$ & - & Battery efficiency\\
\Capa{EV}& $\mathcal{U}$ & \Rplus & kWh & Battery capacity\\
\Pmax{EV}& $\mathcal{U}$ & \Rplus & kW & Charger maximum power\\
$\boldsymbol\alpha^{\text{EV}}$ & $\mathcal{U}$ & \Rplus & \euro/kWh$^2$ & Discomfort reluctance\\
$E^{\text{Cap,EV}}$ & $\mathcal{U}$ & \Rplus & kWh & Total energy consumption\\\hline

\hline

\multicolumn{5}{l}{\rule{0pt}{2ex}\textbf{Water Boiler}} \\\hline
\rule{0pt}{2.5ex}\Temp{ref,WB} & \UxT & \Rplus & \degC & Reference temperature\\
\Temp{set,WB} & \UxT & \Rplus & \degC & Setpoint temperature\\
\Pow{l,WB} & \UxT & \Rplus & kW & Losses through envelope\\
\Pow{u,WB} & \UxT & \Rplus & kW & Losses from water use\\
\Pow{ref,WB} & \UxT & \Rplus & kW & Power reference\\
$\textbf{t}^{\text{u,WB}}$ & \UxT & \Bool & - & Water usage event\\
\Temp{init,WB} & $\mathcal{U}$ & $\left[0,1\right]$ & - & Initial temperature\\
\Capa{th,WB} & $\mathcal{U}$ & \Rplus & $^\circ$/kWh & Thermal coefficient \\
\Pmax{WB} & $\mathcal{U}$ & \Rplus & kW & Water usage event\\
$\boldsymbol\alpha^{\text{WB}}$ & $\mathcal{U}$ & \Rplus & \euro/$^\circ$h & Discomfort reluctance \\
$E^{\text{Cap,WB}}$ & $\mathcal{U}$ & \Rplus & kWh & Total energy consumption\\\hline

\multicolumn{5}{l}{\rule{0pt}{2.5ex}\textbf{Heat Pump}} \\\hline
\rule{0pt}{2.5ex}\Temp{ref,HP} & \UxT & \Rplus & \degC & Reference temperature\\
\Temp{set,HP} & \UxT & \Rplus & \degC & Setpoint temperature\\
\Pow{l,HP} & \UxT & \Rplus & kW & Power losses through walls\\
\Pow{ref,HP} & \UxT & \Rplus & kW & Power from reference profile\\
\Temp{init,HO} & $\mathcal{U}$ & $\left[0,1\right]$ & - & Initial temperature\\
\Capa{th,HP} & $\mathcal{U}$ & \Rplus & $^\circ$/kWh & Thermal coefficient\\
\Pmax{HP} & $\mathcal{U}$ & \Rplus & kW & Nominal thermic power\\
$\textbf{COP}^{\text{HP}}$ & $\mathcal{U}$& \Rplus & - & Coefficient of performance\\
$\boldsymbol\alpha^{\text{HP}}$ & $\mathcal{U}$ & \Rplus & \euro/$^\circ$h & Discomfort reluctance\\
$E^{\text{Cap,HP}}$ & $\mathcal{U}$ & $\Rplus$ & kWh & Total energy consumption
\end{tabular}
\label{tab:inputs}
\end{table}

\thanksto{\noindent This work is supported by the LECaaS project, the UNLEASH project, and the EFES project.}
\vspace{-1cm}
\section{Introduction}

Renewable Energy Communities (RECs) have been proposed as a way to accelerate the energy transition with a bottom-up approach. Indeed, a REC is a promising paradigm for reducing operational stress in the distribution grid while providing benefits to its users by promoting local energy exchanges, economically advantageous compared to traditional retailer exchanges. To unlock the maximum potential of a residential REC, flexible consumption of end-users must be incentivized \cite{pannaganti2023,ISGT24,askeland2025}. As highlighted by some research work, the REC framework increases the willingness of consumers to be flexible with respect to individual consumers \cite{luzzati2024,dudka2025}. Additionally, the growing penetration of low-carbon technologies in residential loads (electric vehicles, water boilers and heat pumps) increases the needs and possibilities for response to residential demand \cite{mathieu2025} and their potential to maximize utility in the residential sector has already been demonstrated \cite{lopez2024}.


This work deals with the coordinated management of end-users' flexible resources to maximize the valorization of local arbitrage in energy communities. To this end, the authors propose an intuitive sequential approach to identify the equilibrium between the community operator activation actions and the independent flexible capacity offered by members. The mobilization of EC members' flexibility is incentivized by the community operator via a requested volume and reward price pair. Consumers respond rationally to these explicit signals according to their individual capabilities, represented in this work by the following flexible loads: water boilers, heat pumps and electric vehicles. Those are intrinsically controllable without user interactions and represent an important part of household consumption when they are present. Finally, besides sending the incentives to the community participants, the community operator also guarantees a fair distribution of the activated volume through a rule-based approach based on regulatory-compliant Keys of Repartition (KoRs) inspired by the Belgian framework \cite{CWAPE,mustika2022}.


In previous work \cite{ISGT24}, we considered how demand-side flexibility can be harnessed from REC members by rewarding their shifted energy volume and discomfort. A two-step model has been proposed for this purpose, where the first step optimally activates the offered flexibility, while the second step shares the benefits through a rule-based or an optimization-based approach. The work presented in \cite{dukovska21} formulated the coordination of REC members for collective energy consumption and production as a mixed-integer linear problem with coupling constraints. The optimization problem is decentralized and solved using a decomposition method.

Alternating direction method of multipliers (ADMM), for distributed optimisation, has been considered in \cite{lilla20,stephant21}. In \cite{lilla20}, the authors focus on a day-ahead operational planning, and privacy-preserving computation is achieved through ADMM, while \cite{stephant21} dealt with the improvement of REC self-consumption. In the latter work, the equilibrium between the community operator and community members was achieved when the grid imported power was below a predefined threshold, instead of focusing on users preferences satisfaction. 

The collective self-consumption is the focus of \cite{mustika2022}. A two-stage approach, decoupling the operational phase from the settlement of energy allocation, has been proposed. A bilevel model is built in \cite{li21} to solve the scheduling problem considering renewable uncertainties and demand response with an emphasis on the flexibility potential of electric vehicles.


Building on our previous work \cite{ISGT24}, our objective was to quantify the gap in collective cost savings between centralized and decentralized coordination of end-user flexible resources for maximization of collective self-sufficiency and self-consumption in energy communities. Compared to the aforementioned works, we propose an intuitive privacy-preserving approach to solve the decentralized problem that allows for embedding more precise (and potentially non-convex) models of flexible appliances, which would noticeably increase the computational complexity of state-of-the-art bilevel and ADMMs methods.
Therefore, our main contributions are:
\begin{enumerate}
    \item we propose a privacy-preserving intuitive sequential approach to solve the decentralized coordination problem of flexible resources in energy communities;
    \item we implement and compare explicit rule-based flexibility activation mechanisms which take the form of regulatory-compliant Keys of Repartition in the decentralized coordination scheme;
\end{enumerate} 
To model flexible resources, we embed an equivalent storage formulation of four relevant flexible domestic appliances (batteries, electric vehicles, heat pumps, water boilers) in the operational load scheduling problem.

In the remainder of this paper, \cref{sec:Framework} describes the framework of the proposed work and the related key assumptions. Then \cref{sec:Problem} presents the centralized flexibility model with an upgraded formulation of flexible devices compared to \cite{ISGT24} before describing the proposed intuitive sequential approach \cref{sec:Approach} to decentralize the model resolution. After that, \cref{sec:Results} presents the case study and showcases the results. Finally, \cref{sec:Concl} presents the main conclusions.

\section{Framework and key assumptions}\label{sec:Framework}
This work considers a REC entity that is constituted by a group of domestic consumers and prosumers (\textit{i.e.}, PV-equipped consumers) that can either exchange electricity with each other or with their traditional retailers. The organization of the local electricity sharing between participants is put in hands of a Community Operator (CO). Besides ensuring local balance and billing of members baseline exchanges within the community, the CO is also responsible for coordinated flexibility scheduling which is the main focus of this paper.

To run this task, assuming a perfect forecast of generation and baseline consumption, the CO identifies the flexibility needs of the day (in both upward and downward directions) of the community to maximize the value of local arbitrage. Then, it passes this information to members along with a reward price. Based on this volume-price pair, EC members offer flexibility capacity (\textit{i.e.}, based on load-shifting) to the operator. Finally, the CO activates the offered capacity explicitly by applying Keys of Repartition.

Hereafter, the different flexible loads considered in this study are presented in detail, as well as the Keys of Repartition compared for the fair activation of flexibility within the REC.

\subsection{Flexible load models}
\begin{figure}[htbp]
    \centering
    \begin{subfigure}[b]{0.1175\textwidth}
        \centering
        \includegraphics{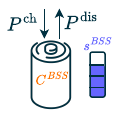}
        \caption{BSS model}
        \label{fig:BSS}
    \end{subfigure}
    \begin{subfigure}[b]{0.1175\textwidth}
        \centering
        \includegraphics{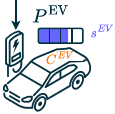}
        \caption{EV model}
        \label{fig:EV}
    \end{subfigure}
    \begin{subfigure}[b]{0.1175\textwidth}
        \centering
        \includegraphics{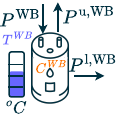}
        \caption{WB model}
        \label{fig:WB}
    \end{subfigure}
    \begin{subfigure}[b]{0.1175\textwidth}
        \centering
        \includegraphics{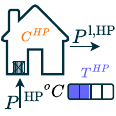}
        \caption{HP model}
        \label{fig:HP}
    \end{subfigure}
    \caption{\centering Equivalent battery models of the four domestic flexible equipments considered.}
    \label{fig:models}
\end{figure}

With the growing electrification of energy demand, new opportunities to unlock flexibility in the consumption profiles of end-users emerge. At the residential level, three main devices have been identified and considered in this work, namely, electric vehicles (EVs), water boilers (WBs) and heat pumps (HPs). The first one, once plugged, allows adaptation of the charging strategy to the needs of the owner, while for the other two, the flexibility comes from the elasticity of the thermal demand as some end-users may have a comfort temperature range and not a specific desired value. In this work, all three flexible load types (or the related demand) are modeled similarly to a battery storage system (BSS) as depicted in Fig.\ref{fig:models}. It means that each one has a state variable (i.e., EV's battery state-of-charge, water temperature of the boiler and indoor air temperature), incoming/outgoing electrical power flows, and a capacity that converts the power exchanges to a change of state (in kWh or in \degC).
\subsection{Repartition keys}\label{sec:keys}
Regarding the activation of flexibility, a rule-based approach is adopted by the community operator. These rules are represented by Keys of Repartition that are usually applied to share the local energy available among members in some countries (\textit{e.g.}, Wallonia (BE) or France). The general KoR mechanism for flexibility activation can be written as:
\begin{align}
    &\Pow{act}_{u,t} = \min(k_{u,t} \Pow{cap}_{0,t}, \Pow{cap}_{u,t}) & \forallut
\end{align}
where $\Pow{cap}_{u,t}$ is the flexible power capacity offered by each member, $\Pow{cap}_{0,t}$ is the requested capacity by the CO, $\Pow{act}_{u,t}$ is the activated flexible power of each member and $k_{u,t}$ is the value of the key of repartition for each individual.

Among the existing KoRs \cite{CWAPE,mustika2022}, three specific ones (\textit{i.e.}, equal, proportional, and cascade) are considered and compared in terms of benefits distribution between flexible members. Their principles are defined as follows: 
\subsubsection{Equal Key}
This key is the simplest one as it splits the requested flexible volume equally between the participants who offer individual flexible capacity.
\begin{equation}
    k\ut = \begin{cases}
        \frac{1}{|\mathcal{U}^{flex}_t|} & \forall u \in \mathcal{U}^{flex}_t,\forallt,\\
        0, & \text{otherwise.}
    \end{cases}
\end{equation}
with $\mathcal{U}^{flex}_t=\{u\in\mathcal{U}|\Pow{cap}\ut>0\}$ the set of REC participants who offered flexible capacity during period $t$.

\subsubsection{Prorate Key}\label{EnergyProrateKey}
This key activates flexibility volumes of each member based on its offered capacity, $\Pow{cap}\ut$,  with respect to the overall capacity available within the REC.
\begin{equation}
    k\ut = \frac{\Pow{cap}\ut}{\sum_{\uinU} \Pow{cap}\ut} \quad  \forallu,\forallt.
\end{equation}

\subsubsection{Cascade Key}
This key consists of an iterative application of the \textit{Equal Key} in order to maximize the activated volume and match the requested one. At each iteration, the set of flexibility providers $\mathcal{U}^{flex}_t$  is recomputed based on the remaining non-activated capacity of each member after the previous activation rounds. The process stops when there remains no offered capacity or the community operator's request is met. The practical application at each time period, $\tinT$, of the \textit{Cascade Key} follows Algorithm \ref{alg:cascade}.
\begin{algorithm}[ht]
\begin{small}
\SetAlgoLined
\caption{Cascade Key principle}\label{alg:cascade}
\SetKwInOut{Input}{Input}
\Input{$\Pow{cap}\ut,\Pow{cap}_{0,t}$}
\SetKwFor{ForEach}{foreach}{do}{end}
\text{Initialize remaining individual and requested flexible capacities:}\\
$\Pow{cap,rem}\ut = \Pow{cap}\ut \quad \forallu$;\\
$\Pow{cap,rem}_{0,t} = \Pow{cap}_{0,t}$;\\
\text{Initialize individual activated volumes:}\\
$\Pow{act}\ut = 0 \quad \forallu$;\\ 
\While{$\sum\limits_{\uinU}\Pow{cap,rem}\ut > 0$ \textbf{and} $\Pow{cap,rem}_{0,t}>0$}{
\text{Compute the number of remaining flexibility providers:}\\
$\mathcal{U}^{flex}_{t}=\{\uinU|\Pow{cap,rem}\ut>0\}$;\\
\text{Apply the Equal Key to activate flexibility:}
$\Pow{act}\ut = \max(\Pow{cap}\ut,\Pow{act}\ut+\Pow{cap,rem}_{0,t}/|\mathcal{U}^{flex}_t|) \quad \forallu$;\\
\text{Update remaining individual and requested flexible capacities:}\\
$\Pow{cap,rem}\ut = \Pow{cap}\ut - \Pow{act}\ut \quad \forallu$;\\
$\Pow{cap,rem}_{0,t} = \Pow{cap}_{0,t} - \sum\limits_{\uinU}\Pow{act}\ut$;\\
}
\end{small}
\end{algorithm}

\section{Centralized Flexibility Activation Problem}\label{sec:Problem}
Building upon previous work \cite{ISGT24}, this section describes the mathematical formulation of the day-ahead coordinated and centralized EC optimal operational planning problem with state-space modeling of the three flexible devices considered (\textit{i.e.}, EVs, HPs and WBs). The objective is to minimize the REC electricity bill by scheduling the use of controllable loads to take advantage of local PV production while accounting for the discomfort caused by load-shifting. The complete problem, denoted as \textit{ECFlex}, is written as:
\allowdisplaybreaks{
\begin{subequations}\label{PB:model_indices}
\begin{flalign}
&\min_{\boldsymbol\Omega}\hspace{1cm}\textstyle\sum_{\uinU}(\bill{}_u  + \discomf{flex}_u)\\
&\text{subject to :} &\notag\\
&\textstyle\bill{}_u = \timestep\sum_{\tinT}(\price{i,ret}\ut\importGrid\ut - \price{e,ret}\ut\exportGrid\ut &\notag\\
&\hspace{2.55cm} + \OPFee(\importCom\ut+\exportCom\ut) ), \hspace{-0.38cm}& \forallu \label{eq:Bill_indices}\\
& \textstyle \discomf{flex}_u = \sum_{\tinT}(\discomf{EV}\ut + \discomf{WB}\ut + \discomf{HP}\ut), & \forallu \label{eq:Disc_indices}\\
& \Pinj\ut = \PPV\ut + \discharge\ut - \fix\ut  - \flex\ut - \charge\ut, \hspace{-0.4cm}&\forallut \label{eq:Pphys_indices}\\
& \Pinj\ut = \exportGrid\ut + \exportCom\ut -\importGrid\ut - \importCom\ut, & \forallut \label{eq:Pvirt_indices}\\
& \textstyle\sum_{\uinU}\exportCom\ut = \sum_{\uinU}\importCom\ut, & \forallt \label{eq:Combal_indices}\\
& \charge\ut \leq \Pmax{BSS}_u, & \forallut \label{eq:BSSChamax_indices}\\
& \discharge\ut \leq \Pmax{BSS}_u, & \forallut \label{eq:BSSDismax_indices}\\
& \SoCmin{BSS}_{u} \leq \SoC{BSS}\ut \leq \SoCmax{BSS}_{u}, &\forallut\label{eq:BSSlim_indices}\\
& \SoC{BSS}\ut = \SoC{prev,BSS}_{u,t} +  \timestep (\eff{BSS}_u\charge\ut & \notag\\
& \hspace{2.7cm}- \discharge\ut/\eff{BSS}_u)/\Capa{BSS}_u, \hspace{-0.4cm}&\forallut\label{eq:BSSsoc_indices}\\
& \SoC{BSS}_{u,T} = \SoC{init,BSS}_u, & \forallu \label{eq:BSSEnd_indices}\\
& \flex\ut = \Pow{EV}\ut + \Pow{WB}\ut + \Pow{HP}\ut, & \forallut \label{eq:Pflex_indices}\\
& \Pow{EV}\ut \leq \text{a}\ut^{\text{EV}} \Pmax{EV}_u, & \forallu \label{eq:EVmax_indices}\\
& \SoC{EV}\ut = t^{\text{arr,EV}}\ut\SoC{arr, EV}\ut + (1-t^{\text{arr,EV}}\ut)\SoC{EV,prev}_{u,t} \hspace{-0.40cm}&\notag\\
& \hspace{2.75cm} + \timestep \eff{EV}_u \Pow{EV}\ut/\Capa{EV}_u, \hspace{-0.38cm}& \forallut \label{eq:EVsoc_indices}\\
& \text{t}\ut^{\text{dep,EV}} \SoC{ref,EV}_{u,t} \leq \SoC{EV}\ut \leq 1, & \forallut \label{eq:EVtarget_indices}\\
& \textstyle \sum_{\tinT}\Pow{EV}\ut = \sum_{\tinT}\Pow{ref,EV}\ut, & \forallu\label{eq:EVtotal_indices}\\
&\discomf{EV}\ut \geq \alpha^{\text{EV}}_u \left( \SoC{ref,EV}\ut - \SoC{EV}\ut\right), &\forallut \label{eq:EVdisc_indices}\\
& \Pow{WB}\ut \leq \Pmax{WB}_u & \forallut \label{eq:WBmax_indices}\\
& \Temp{WB}\ut = \Temp{prev,WB}_{u,t} + \timestep (\Pow{WB}\ut - \Pow{u, WB}\ut& \notag\\
& \hspace{2.8cm} - \Pow{l,WB}\ut)/\Capa{th,WB}_u, \hspace{-0.38cm}&\forallut \label{eq:WBtemp_indices}\\
& \Tempmax{WB}\ut \geq \Temp{WB}\ut \geq \text{t}^{\text{u,WB}}\ut \Temp{lim,WB}\ut, & \forallut \label{eq:WBtarget_indices}\\
& \textstyle \sum_{\tinT}\Pow{WB}\ut = \textstyle \sum_{\tinT}\Pow{ref,WB}\ut, & \forallu \label{eq:WBtotal_indices}\\
& \discomf{WB}\ut \geq \alpha^{\text{WB}}_u (\Temp{lim,WB}\ut  - \Temp{WB}\ut), &\forallut\label{eq:WBdisc_indices}\\
& \Pow{HP}\ut \leq \Pmax{HP}_u, & \forallut \label{eq:HPmax_indices}\\
& \Temp{HP}\ut = \Temp{prev,HP}_{u,t} +\timestep(COP^\text{HP}_u\Pow{HP}_u  & \notag\\
& \hspace{2.8cm}- \Pow{l,HP}\ut)/\Capa{th,HP}_u, \hspace{-0.38cm}&\forallut \label{eq:HPtemp_indices}\\
& \textstyle \sum_{\tinT}\Pow{HP}\ut = \textstyle \sum_{\tinT}\Pow{ref,HP}\ut, & \forallu \label{eq:HPtotal_indices}\\
& \discomf{HP}\ut \geq \alpha^{\text{HP}}_u (\Temp{lim,HP}\ut - \Temp{HP}\ut). &\forallut\label{eq:HPdisc_indices}
\end{flalign}
\end{subequations}
}
where $\boldsymbol\Omega =[\charge,\discharge,\Pow{EV},\Pow{WB},\Pow{HP}, \importCom, \exportCom]$ are the decision variables of the community operator based on participants' inputs. Regarding the objective function, \eqref{eq:Bill_indices} describes the electricity bill components: the cost for purchasing complementary electricity from an external retailer, revenues for selling remaining production surplus, and the operation cost of the community exchanges (\textit{e.g}., to cover CO workload compensation or grid fees). Equation \eqref{eq:Disc_indices} indicates that discomfort can arise from the flexible use of the three considered assets, as made explicit later. \eqref{eq:Pphys_indices}-\eqref{eq:Combal_indices} represent respectively, the physical power balance at the point of common coupling of EC members with the network, the economical balance between local and external exchanges, and the community-level balance between purchased and sold volumes. Regarding battery storage systems, \eqref{eq:BSSChamax_indices}-\eqref{eq:BSSlim_indices} imposes physical bounds to the controlled power and state-of-charge variables while \eqref{eq:BSSsoc_indices} represents the dynamics of the battery state-of-charge over time with $\SoC{prev,BSS}\ut$ being equal to $\SoC{BSS}_{u,t-1}$ except at time slot $t=1$ for which an initial level, $\SoC{init,BSS}_u$ is considered. The latter level is imposed to be recovered by the end of the optimization period through \eqref{eq:BSSEnd_indices}. Finally, the flexible devices models, presented in Fig.\ref{fig:models}, are similarly translated in mathematical constraints and the power consumption of all devices are aggregated under the flexible power variable in \eqref{eq:Pflex_indices}. First, \eqref{eq:EVmax_indices}, \eqref{eq:WBmax_indices} and \eqref{eq:HPmax_indices} set the upper power bounds based on devices ratings (and EVs presence at the charging point). \eqref{eq:EVsoc_indices}, \eqref{eq:WBtemp_indices} and \eqref{eq:HPtemp_indices} model the evolution of state variables that are respectively for EVs, WBs and HPs, the state-of-charge of the vehicle's battery, the hot water temperature and the ambient air temperature. Each of those variables evolution is affected by the consumed electrical power, the final energy use and dissipative losses. Similarly to BSS, the $\text{prev}$ superscript correspond to the state at $t-1$ except for $t=1$ for which an initial reference state (\textit{i.e.}, $\SoC{ref,EV}_{u,1}$ for EVs, $\Temp{lim,WB}_{u,1}$ for WBs and $\Temp{lim,HP}_{u,1}$ for HPs) is imposed. Additionally, in \eqref{eq:EVsoc_indices} the state-of-charge of the vehicle is updated at the end of a trip, when arriving at the charging station. \eqref{eq:EVtarget_indices} and \eqref{eq:WBtarget_indices} respectively ensures EVs reached the targeted charge level at departure time and hot water temperature is at the requested temperature when used. Finally, \eqref{eq:EVtotal_indices}, \eqref{eq:WBtotal_indices} and \eqref{eq:HPtotal_indices} guarantee a constant daily load with respect to reference as only load shifting is allowed and \eqref{eq:EVdisc_indices}, \eqref{eq:WBdisc_indices} and \eqref{eq:HPdisc_indices} defines linear discomfort reaction when the targeted state is not met.


\subsection*{Additional benchmark models}
In order to assess the local arbitrage benefits arising from optimal flexible devices coordination, benchmark frameworks must be defined. In this work, three additional models are investigated that derive from the problem \eqref{PB:model_indices} by adding specific additional constraints.
\subsubsection{Baseline problem (SoloFix)}
This constitutes the reference case and is associated to the idea of solitary end-user models (\textit{i.e.}, no local sharing activities) without flexibility concerns (\textit{i.e.}, fixed power consumption profile). This typical scenario serves as the baseline for the subsequent models. It is obtained by adding the following constraints to problem \eqref{PB:model_indices}:
\begin{subequations}
\begin{flalign}
    &\exportCom\ut = \importCom\ut = 0 & \forallut \label{eq:NoCom}\\
    & \flex\ut = \Pow{ref,EV}\ut + \Pow{ref,WB}\ut + \Pow{ref,HP}\ut & \forallut \label{eq:NoFlex}
\end{flalign}
\end{subequations}

\subsubsection{Individual self-consumption problem (SoloFlex)}
Thereafter, the second benchmark introduces participants' flexibility potential in order to take advantage of individual local renewable generation and maximize its self-consumption. Compared to the baseline problem, in this case only the constraint \eqref{eq:NoCom} is added to problem \eqref{PB:model_indices}.

\subsubsection{Non-flexible community problem (ECFix)}
Lastly, a collective problem can be formulated without addressing the management of flexible resources except for the batteries. From a mathematical point of view, this case adapts \eqref{PB:model_indices} by adding the non-flexibility constraint \eqref{eq:NoFlex} but keeping local electricity exchanges allowed. 

\subsubsection{Prioritization of individual self-consumption (\textit{ECFlex'})}

Usually used as a comparison benchmark, like in \cite{ISGT24}, the \textit{SoloFlex} model may also be seen as preliminary actions before coordinated flexibility planning. This aspect is studied in this work by changing the reference power consumption profile of controllable appliances by the output values of $\Pow{EV}, \Pow{WB} \& \Pow{HP}$ from the individual self-consumption problem. In this new framework, individual self-consumption is prioritized by design before the EC coordinated flexibility actions are taken\footnote{Note that state variables reference values are not adapted in the discomfort functions of problem \eqref{PB:model_indices}, meaning that the discomfort created by individual flexibility actions is passed to the coordinated problem.}. The use of these new reference powers in further models will be indicated by a quotation mark ($\Powprim{ref,EV}\ut, \Powprim{ref,WB}\ut, \Powprim{ref,HP}\ut$).
\section{Proposed decentralized resolution approach}\label{sec:Approach}
If the centralized approach offers collective optimality guarantees, as presented in \cite{ISGT24}, it raises privacy concern and limits the active implication of end-users in the flexibility provision. The latter may also negatively affect the fairness of benefits distribution, as it may prioritize some individuals based on the resources they have and not on their willingness to participate. To alleviate this issue, the authors describe in the following the decentralized resolution approach of the flexibility activation problem presented in \cref{sec:Problem}.

First, let us split the set of decision variables with
$\boldsymbol\Omega_u  =[\charge,\discharge,\Pow{EV},\Pow{WB},\Pow{HP},\Pow{cap,+},\Pow{cap,-}]$ being the individual decision variables of REC members while $\boldsymbol\Omega_{\text{CO}} = [\importCom, \exportCom,\Pmax{act,+}\ut,\Pmax{act,-}\ut]$ are the decision variables of the community operator. 
Then, let us define $\importGrid_{0,t}$, $\exportGrid_{0,t}$, as the net import and net export, respectively, at the point of common coupling (PCC) of the REC, which represent the collective external exchanges obtained after resolution of the \textit{Non-flexible community problem}. They serve as initial downward and upward flexibility requests (\textit{i.e.}, capacity limits) of the CO at each timestep as presented in Algorithm \ref{alg:Decentralized}. The third input is the total flexible device reference power $\Pow{ref}\ut = \Pow{flex}\ut + \Pow{cha}\ut - \Pow{dis}\ut $.  

Once the CO has identified the initial flexibility needs of the REC, flexible capacity will be offered by end-users and validated (or not) by the operator iteratively until there is no more flexibility request at the collective scale (\textit{i.e.}, no more local generation to take advantage of) or no member is offering additional flexible capacity.

At each iteration, there is a two-way communication between REC members and the CO for the activation of flexibility that follows these three steps:
\begin{enumerate}
    \item Based on the community level flexibility request, each member tries to maximize its revenues from the provision of flexibility by submitting capacity offers to the CO, knowing they can earn $\price{act} = \price{i,ret} - \price{e,ret} - 2\OPFee$ for each displaced kWh. These individual problems are constrained by the set of operating constraints of flexible resources and the following additional ones: 
    \allowdisplaybreaks{
    \begin{subequations}
        \begin{flalign}
        &\Pow{cap,+/-}\ut \leq \Pmax{cap,+/-}_{0,t}, &\forallut,\label{eq:Pcap1}\\
        &\textstyle \sum_{\tinT} \Pow{cap,+}\ut = \sum_{\tinT} \Pow{cap,-}\ut, &\forallu, \label{eq:Pcap2}\\
        &\Pow{flex}\ut = \Pow{ref}\ut + \Pow{cap,+}\ut - \Pow{cap,-}\ut,& \forallut. \label{eq:Pcap3} 
        \end{flalign}
        \end{subequations}}
\item The resulting capacity offers obtained by the CO are not aware of other members' decisions. They are provided as if each end-user were the sole member of the REC. Therefore, to ensure a fair contribution to the flexibility activation for everyone, the operator refines individual activation bounds, $\Pmax{act,+/-}_{u,t}$, by distributing the flexibility request among the members through the use of a Keys of Repartition mechanism as defined in \cref{sec:keys}.
\item Finally, as the newly computed individual bounds for each timestep may not necessarily ensure constraint \eqref{eq:Pcap2}, consumers solve again the maximization of flexibility revenues problem with the updated bounds as follows:
\begin{flalign}
     &\Pow{cap,+/-}\ut \leq \Pmax{act,+/-}_{u,t}, & \forallut \label{eq:Pact1}.
\end{flalign}
The resulting flexible resource usage decisions are effectively activated and set the new reference consumption for the next iteration. From the CO perspective, the EC level request is updated to account for the already activated flexibility. 
\end{enumerate}

\begin{algorithm} [t]
\begin{small}
\SetAlgoLined
\caption{Proposed decentralized resolution algorithm \textit{ECFlexIt}}\label{alg:Decentralized}
\SetKwInOut{Input}{Input}
\Input{$\importGrid_{0}$, $\exportGrid_{0}$, $\Pow{ref}$}
\SetKwFor{ForEach}{foreach}{do}{end}
\hangindent=0.5\skiptext\hangafter=1 Set the initial CO flexibility requests as upper bounds for EC members offers:\\ 
$\Pmax{cap,+}_{0,t}, \Delta\Pmax{cap,+} \gets \exportGrid_{0,t} \quad \forallut$;\\
$\Pmax{cap,-}_{0,t}, \Delta\Pmax{cap,-} \gets \importGrid_{0,t} \quad \forallut$;\\
\While{$\Pmax{cap,+/-}_{0,t} > 0$ \textbf{and} $\Delta\Pmax{cap,+/-}>0$}{
\ForEach{$u \in \mathcal{U}$}{
\text{Individuals solve flexibility capacity problem:}\\
\hangindent=0.5\skiptext\hangafter=1 $\Pow{cap,+/-}\ut \gets \Big\{\argmax\limits_{\Omega_u} \hspace{-0.1cm}\sum\limits_{\tinT} \Pow{+,cap}\ut \price{act}\timestep - \discomf{flex}_u,$ \\
$\hspace{3cm}\text{s.t. } \eqref{eq:BSSChamax_indices}-\eqref{eq:HPdisc_indices}, \eqref{eq:Pcap1}-\eqref{eq:Pcap3}\Big\}$;\\}
\hangindent=0.5\skiptext\hangafter=1 The CO applies a key of repartition to define individual flexibility activation bounds:\\
$\Pmax{act,+/-}\ut \gets KoR(\Pow{cap,+/-}) \quad \forallut$;\\ 
\ForEach{$u \in \mathcal{U}$}{
\hangindent=0.5\skiptext\hangafter=1 Each member activates its flexibility based on the new individual limits:\\   
\hangindent=0.5\skiptext\hangafter=1 $\Pow{act,+/-}\ut = \Pow{cap,+/-}\ut \gets \Big\{\argmax\limits_{\Omega_u} \hspace{-0.1cm}\sum\limits_{\tinT} \Pow{+,cap}\ut \price{act}\timestep - \discomf{flex}_u,$\\
$\hspace{2cm}\text{s.t. } \eqref{eq:BSSChamax_indices}-\eqref{eq:HPdisc_indices}, \eqref{eq:Pcap2}-\eqref{eq:Pcap3}, \eqref{eq:Pact1}\Big\}$;\\
\hangindent=0.5\skiptext\hangafter=1 The activated flexibility is taken into account for the next iteration by updating the reference load profiles:
$\Pow{ref}\ut \gets \Pow{ref}\ut + \Pow{act,+}\ut - \Pow{act,-}\ut$;}
\hangindent=0.5\skiptext\hangafter=1 The CO updates its EC level flexibility requests and keeps track of the activated volume:\\
$\Pmax{cap,+/-}_{0,t} \gets \Pmax{cap,+/-}_{0,t} - \sum_{\uinU} \Pow{act,+/-}\ut \quad \forallu,\forallt$;\\
$\Delta\Pmax{cap,+/-} \gets \textstyle\sum_{\uinU}\sum_{\tinT} \Pow{act,+/-}\ut $;
}
\end{small}
\end{algorithm}


\section{Case Study and Results}\label{sec:Results}
This section provides detailed results for a full year of simulation and compares the economic and energy usage performances of the proposed decentralized resolution approach to the benchmark models presented in \cref{sec:Problem} and the centralized approach developed in \cite{ISGT24}. 

The considered energy community is composed of 20 domestic end-users whose non-flexible and flexible components are generated using a stochastic residential load profile generator \textit{Resflex} \cite{stegen2025residential}. The penetration levels of the four flexible assets considered in the 20 households are 70\% for WBs, 60\% for EVs, 50\% for HPs, and 25\% for BSS. Among the members, 15 also have PV installations ranging from 2 to 20 kW$^{\text{peak}}$, for a total of 147 kW$^{\text{peak}}$ at the community level. Regarding the electricity tariffs, only flat prices are considered with an import price (including retail and grid fees) of 0.4 €/kWh, an export price of 0.1 €/kWh and a community fee of 0.01 €/kWh charged by the CO. Hence, the reward value for each kWh of flexibility activated rises to 0.28 €/kWh. 

The community members have a day-ahead planning for the usage of their flexible appliances (EV charging and HP/WB temperature setpoints and predicted losses) and the model is simulated for one full year with sequential solving of daily problems.
Finally, \textit{ex-post} power flow analyses are run for the different scenarios studied, assuming that REC participants are connected to a benchmark low voltage distribution grid \cite{dickert}.


\begin{table}[htbp]
\centering
\caption{Summary of results for selected scenarios.}
\label{tab:discomfort}
\renewcommand{\arraystretch}{0.4}
\begin{tabular}{l|rrrrr}
 & \textbf{SoloFlex} & \textbf{ECFix}   & \textbf{ECFlex}   & \textbf{ECFlexIt}   & \textbf{ECFlexIt'}  \\\hline
 \rule{0pt}{2ex}[\euro] & & & \\
  \bill{}         & 6,549   & 6,831   & 5,243   & 5,576    & 5,412    \\
  \discomf{EV}    & 79      & 0       & 188     & 36       & 802      \\
  \discomf{WB}    & 0       & 0       & 0       & 0        & 0        \\
  \discomf{HP}    & 0       & 0       & 0       & 0        & 0        \\\hline
\rule{0pt}{2ex}[MWh]& & & \\
 $E^{\text{act}}$  & 40.6   & 0       & 44.9 & 23.7    & 12.7       \\
 $E^{\text{act,EV}}$     & 2.3    & 0       & 4.1  & 0.8     & 0.6       \\
 $E^{\text{act,WB}}$     & 27.4   & 0       & 30.7 & 17.1    & 8.5        \\
 $E^{\text{act,HP}}$     & 10.9  & 0       & 10.2 & 5.9     & 3.6\\
 $E^{\text{dis,BSS}}$     & 7.3  & 10.2       & 7.4 & 10.1     & 8.6 
\end{tabular}
\end{table}




\subsection{Preserving privacy does not imply high benefits loss}

\begin{figure}
    \centering
    \includegraphics[width=\linewidth]{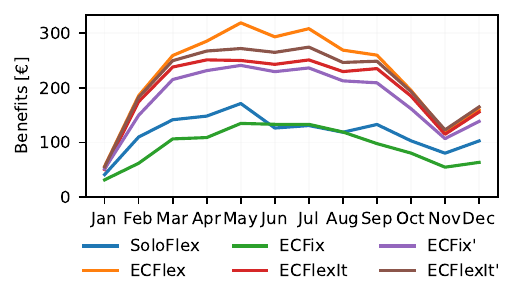}
    \caption{Total benefits over the SoloFix model}
    \label{fig:bills}
\end{figure}

One of the key findings of this study is that the coordinated decentralized iterative approach described in Section~\ref{sec:Approach} converges towards the centralized methodology presented in previous work~\cite{ISGT24}. As illustrated in Figure~\ref{fig:bills}, the proposed \textit{ECFlex} and \textit{ECFlexIt'} (\textit{i.e.}, with and without individual prioritization and with an Equal Key) models provides a close, though suboptimal approximation of the case in which the decision-making process is fully coordinated by the CO. Indeed, a residual billing inefficiency remains, as reported in Table~\ref{tab:discomfort}, with a respective 6.35\% and 3.22\% deviation observed in the overall community cost compared to the centralized benchmark. Nonetheless, this deviation represents the trade-off required to enhance user privacy and decentralize the decision-making process to community members, thereby ensuring individual preferences are more explicitly considered compared to the fully centralized case where the operator optimizes the overall REC outcome.

In addition, one should take a look at the other component of the objective function, that is, the discomfort generated by the activation of flexibility resources. In this perspective, the first interesting outcome observed in Table~\ref{tab:discomfort} is that water boilers and heat pumps have the potential to activate a significant amount of flexibility (up to 30.7\% for WBs and 10.2\% for HPs) without causing any discomfort, despite $\alpha^{HP}$ and $\alpha^{WB}$ equal to 1. This is made possible as only lower temperatures than the reference ones are penalized. This allows the thermal assets to shift the evening consumption to solar hours to "overheat" the water and air and let it cool down slowly over the end of the day without going below the limit. This emphasizes the flexibility potential at low or negligible costs from other assets than typical BSS, often unique flexible resources considered in this type of study.  

On the other hand, shifting EV power only comes with discomfort costs as it delays full charge and limits anticipated departure potential. They are therefore used in cases where other assets do not have enough potential to take advantage of the overall local production available.

Comparing the different scenarios together, one can also notice that in the decentralized framework, without prior self-consumption, the initial use of storage systems before the activation of other flexibility means limits the action of the latter. The centralized approach, on the contrary, can make the direct arbitrage between the free WBs and HPs flexibility and the imperfect battery systems (i.e., 95 $\%$ efficiency). The prioritization of individual self-consumption allows for applying similar arbitrage at the individual level before coordinating the responses to remaining flexibility requests of the CO.   


\subsection{KoRs have a limited influence on fair benefits repartition}


\begin{figure}
    \centering
    \includegraphics{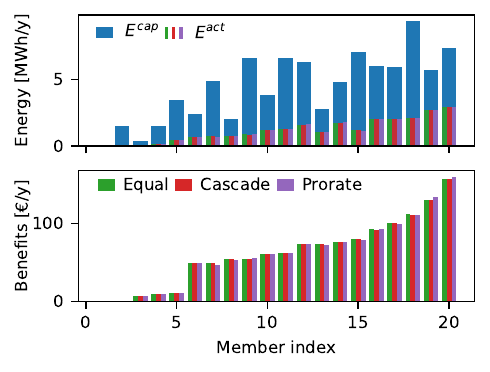}
    \caption{Individualization of the benefits compared to the community with fixed consumption}
    \label{fig:Indiv}
\end{figure}
The proposed decentralized resolution approach has been applied without prioritization for the different Keys of Repartition presented in \cref{sec:keys} to assess their impact on the individual benefits distribution. 

Nevertheless, results displayed in Figure \ref{fig:Indiv} show that there are very small differences in terms of flexible energy activation and consequent benefits for the three different key mechanisms. Indeed, as discussed in the previous results section, the main driving factor for the participation of individuals to the flexibility coordination procedure is the flexible assets ownership. Hence, users with EVs only (User 2 in Fig. \ref{fig:Indiv}) may have an interesting flexible capacity (\textit{i.e.}, the overall consumption of owned flexible assets), $E^{cap}$, but they will tend to offer limited capacity to the operator due to the discomfort that it creates. Meanwhile, REC members combining the two thermal assets (WBs and HPs) (\textit{e.g.}, User 20 in Fig. \ref{fig:Indiv}) can offer a large part of their flexible capacity without a negative counterpart, raising their economic benefits. Regarding heat pumps, it must be pointed out that when they are only considered for heating households, their action is limited during summer periods, while those with WB exhibit the highest potential for local exchanges.    

Nevertheless, the chosen Keys of Repartition have an impact on the convergence of the iterative procedure, as the cascading and proportional activation will tend to reduce the number of required iterations by dispatching all the flexibility offered at each step.


\begin{table}[!htb]
\setlength{\tabcolsep}{4pt}
\centering
\caption{Priorization impact}
\begin{tabular}{l|rlrr|rlrr|}
 & \multicolumn{4}{c|}{With self-priorization} & \multicolumn{4}{c|}{Without priorization} \\
      &   B [k€] & $\Delta$ B   &   J [€] &   $E^{\text{act}}$&   B [k€] & $\Delta$ B   &   J [€] &   $E^{\text{act}}$ \\
\hline
 \multicolumn{1}{l|}{\rule{0pt}{2ex}SoloFix}  &   7.9 &       &    &      &   7.9 &    & 0    &   0    \\
 \multicolumn{1}{l|}{ SoloFlex} &   6.5 & -1.4  & 79  &   40   &&&&   \\
 \multicolumn{1}{l|}{ ECFix}    &   5.7 & -0.8  & 79  &   40   &   6.8 & -1.1 & 0    &   0    \\
 \multicolumn{1}{l|}{ ECFlexIt} &   5.4 & -0.3  & 803 &   53   &   5.6 & -1.2 & 37   &  24 \\
\end{tabular}
\end{table}



\subsection{Flexibility impacts on the grid require explicit incentives}
Regarding grid performances, it is observed that the yearly average indicators improve in scenarios where flexibility activation is enabled. 
Specifically, the total power losses are reduced from 840 kWh (\textit{ECFix}) to 690 kWh (\textit{ECFlex'}) in the considered distribution network, improving the local grid efficiency. On top, the activation of flexibility mechanisms results in approximately 12 MWh less energy exchanged with the upstream grid in both directions (both imported and exported energy). This enhanced self-consumption and self-sufficiency of the community also benefits to the global grid by reducing the power flows through the upstream lines.

\section{Conclusion}\label{sec:Concl}
This work aimed to propose an intuitive resolution approach for the decentralized coordination of flexible resources within a Renewable Energy Community to address the privacy concern of typical centralized methods such as proposed in \cite{ISGT24} without increasing the computational burden induces by state-of-the-art bilevel or ADMMs techniques. On a 20-household case study, results show that one can reach collective benefits close to the one obtained with centralized coordination with only $3.22 \%$ deviation in the best case. Authors also identified thermal loads such as water boilers and heat pumps as promising flexible resources as their smart management can improve the usage without generating any discomfort for the members. Finally, the investigation of several Keys of Repartition mechanism for the activation of offered flexible capacity did not lead to key differences as the individual benefits distribution is mainly driven by the flexible assets ownership. 
\comment{
\section{Model without indices}

\begin{subequations}\label{PB:model2}
\begin{flalign}
\min_{\boldsymbol\Omega}&\hspace{1cm}\onemat{1}{U}\cdot(\bill{}  + \discomf{flex}) \cdot \onemat{T}{1}\\
\text{s.t.  } & \bill{} = \timestep(\diag{\price{i,ret}} \odot \importGrid \label{eq:Bill}\\
& \hspace{2em} - \diag{\price{e,ret}} \odot \exportGrid \notag\\
&\hspace{2em} + \diag{\OPFee}\odot(\importCom+\exportCom) )\notag\\
& \discomf{flex} = \discomf{EV} + \discomf{WB} + \discomf{HP} \label{eq:Disc}\\
\notag\\ 
& \Pinj = \PPV + \discharge - \fix  - \flex - \charge \label{eq:Pphys}\\
& \Pinj = \exportGrid + \exportCom -\importGrid - \importCom \label{eq:Pvirt}\\
& \onemat{1}{U} \cdot \exportCom = \onemat{1}{U} \cdot \importCom \label{eq:Combal}\\
& \flex = \Pow{EV} + \Pow{WB} + \Pow{HP} \label{eq:Pflex}\\
& \PPV \leq \diag{\Capa{PV}}\PPVprof \cdot \onemat{1}{T} \label{eq:PVmax}\\
&\SoC{BSS} = \SoC{prev, BSS} +\label{eq:BSSsoc}\\
&\hspace{3em} \timestep (\diag{\eff{BSS}}\charge - \discharge/\diag{\eff{BSS}})/\Capa{BSS}\notag\\
&\notag \\
& \Pow{EV} \leq \textbf{a}^{\text{EV}} \odot \Pmax{EV}\cdot\onemat{1}{T} \label{eq:EVmax}\\
& \SoC{EV} \geq \textbf{t}^{\text{dep,EV}} \odot \bar{s}^{\text{EV}} \cdot \onemat{1}{T}\\
& \hspace{2.5em} + \timestep \diag{\eff{EV}} \Pow{EV}/\Capa{EV} \notag\\
& \SoC{EV} = \SoC{arr, EV} + (\onemat{U}{T}-\textbf{t}^{\text{arr,EV}}) \odot \SoC{prev, EV} \label{eq:EVsoc}\\
& \Pow{EV}\cdot\onemat{T}{1} = \Pow{ref,EV}\cdot\onemat{T}{1}\\
&\discomf{EV} \geq \diag{\boldsymbol\alpha^{\text{EV}}} \left( \SoC{EV} - \SoC{ref,EV}\right) \label{eq:EVdisc}\\
\notag\\
& \Pow{WB} \leq \Pmax{WB} \cdot \onemat{1}{T} \label{eq:WBmax}\\
& \Temp{WB} = \Temp{prev,WB} + \\
& \hspace{2.5em}\diag{\Capa{th, WB}} \timestep (\Pow{WB} - \Pow{u, WB} - \Pow{l,WB})\notag\\
& \Temp{WB} \geq \textbf{t}^{\text{u,WB}} \odot \min\{\Temp{ref,WB},\Temp{set,WB}\}\\
& \Pow{WB}\cdot\onemat{T}{1} = \Pow{ref,WB}\cdot\onemat{T}{1}\\
& \discomf{WB} \geq \diag{\boldsymbol\alpha^{\text{WB}}} (\min\{\Temp{ref,WB},\Temp{set,WB}\} - \Temp{WB})\label{eq:WBdisc}\\
\notag\\
& \Pow{HP} \leq \Pmax{HP} \cdot \onemat{1}{T} \label{eq:HPmax}\\
& \Temp{HP} = \Temp{prev,HP} + \\
& \hspace{2.5em}\Capa{th,HP} \timestep(\diag{COP^\text{HP}}\Pow{HP} - \Pow{l,HP})\notag\\
& \Pow{HP}\cdot\onemat{T}{1} = \Pow{ref,HP}\cdot\onemat{T}{1}\\
& \discomf{HP} \geq \diag{\boldsymbol\alpha^{\text{HP}}} (\min\{\Temp{ref,HP},\Temp{set,HP}\} - \Temp{HP}) \label{eq:HPdisc}
\end{flalign}
\end{subequations}
where, $\diag{\textbf{x}}$ is the diagonal matrix constructed with the components of $\textbf{x}$, $\textbf{x}^{\text{\_,prev}}$ is a matrix whose first column is $\textbf{x}^{\text{\_,init}}$ concatenated with the columns $\left[1,T-1\right]$ of the matrix $\textbf{x}^{\text{\_}}$, $\odot$ is the Hadamart product operator, \onemat{n}{m} matrix of ones of size $n \times m$.
\begin{alignat}{2}
\!\min\hspace{1cm} & \text{Individual bills and discomfort} & \notag\eqref{PB:model2}\\
\text{subject to} \hspace{.5cm} & \textit{Cost equations} \hspace{.2cm} \eqref{eq:Bill}-\eqref{eq:Disc} ,\notag\\
&\textit{Power exchange constraints} \hspace{.2cm} \eqref{eq:Pphys}-\eqref{eq:Pflex}, \notag\\
&\textit{PV and BSS model} \hspace{.2cm} \eqref{eq:PVmax}-\eqref{eq:BSSsoc}, \notag\\
&\textit{EV model} \hspace{.2cm} \eqref{eq:EVmax}-\eqref{eq:EVdisc},\notag\\
&\textit{WB model} \hspace{.2cm} \eqref{eq:WBmax}-\eqref{eq:WBdisc},\notag\\
&\textit{HP model} \hspace{.2cm} \eqref{eq:HPmax}-\eqref{eq:HPdisc}.\notag
\end{alignat}

\newcommand{\onemat}[2]{\ensuremath{\mathbf{1}_{#1\times#2}}}

$t^{arr}_{e_u}$ and $t^{dep}_{e_u}$, and the arrival and departure state of charge, $s^{arr,EV}_{e_u}$ and $s^{dep,EV}_{e_u}$.  
\begin{align}
s^{EV}_{u,t^{arr}_{u}} = &s^{arr,EV}_{u} + \timestep P^{EV}_{u,t^{arr}_{u}} &&\forallu, \label{EV:Arr} \\
s^{EV}_{u,t} =& s^{EV}_{u,t-1} + \timestep P^{EV}_{u,t} && \forallu, \forall t\in]t^{arr}_{u},t^{dep}_{u}],
\label{EV:SOC}\\
s^{EV}_{u,t^{dep}_{u}} = &s^{dep,EV}_{u} && \forallu, \label{EV:Dep} \\
 0 \leq &P^{EV}_{u,t} \leq \overline{P}^{EV}_{u}\delta^{EV}_{u,t} && \forallu, \forallt, \label{EV:Pmax}  \\
 \underline{S}^{EV}_{u} \leq &s^{EV}_{u,t} \leq \overline{S}^{EV}_{u} && \forallu, \forallt. \label{EV:SOCmax} 
\end{align}
with $s^{EV}_{e_u,t}$ the electric vehicle state-of-charge, $P^{EV}_{u,t}$ the charging power of the electric vehicle that is capped by the charger rating when the vehicle is plugged and is capped to $0$ when the vehicle is unplugged. The presence or not of the vehicle at the charging station is represented by a binary vector $\delta^{EV}_{u,t}$.

\begin{align}
    &  T^{in}_{t+1} = T^{in}_t + \timestep k_{th} * (P^{HP,th}_{u,t} - P^{loss}_{u,t}) & \forallu, \forallt \\
    & \underline{T}^{in}_u \leq T^{in}_t \leq \bar{T}^{in}_u & \forallu, \forallt \label{HP:Tbound}\\ 
    & 0\leq P^{HP,el}_{u,t} \leq \bar{P}^{HP}_{u} & \forallu, \forallt \\
    & P^{HP,th}_{u,t} = COP_{u,t} * P^{HP,el}_{u,t} & \forallu, \forallt\label{HP:COP}
\end{align}
with $T^{in}\ut$, the inner temperature of buildings, $k^{th}_u =  \textstyle \frac{1}{m_{air}*c_{p,air}*ICF}$ the thermal coefficient, and $P^{loss}_t = P^{exfiltration}_t - P^{irradiation}_t - P^{conduction}_t$ the thermal loss of the house computed at time t. 
$P^{H}_{h_u,t}$ us the power consumption of thermal loads and $P^{ref,H}_{h_u,t}$ the base consumption profile of those loads.

\subsubsection{Discomfort functions}
The overall discomfort created by a flexible behavior to one member includes thermal discomfort and electric vehicle late state-of-charge fulfill:
\begin{align}
    J^{cap,tot}_u=& \sum_{e_u \in \mathcal{E}_u}J^{cap,EV}_{e_u} + J^{cap,TH}_{u}&&\forallu\label{Comfo:Tot}\\
    J^{cap,EV}_{e_u}=& \alpha_u\sum_{t \in \mathcal{T}} (s^{EV,ref}_{e_u,t} - s^{EV}_{e_u,t}) && \forall e_u \in \mathcal{E}_u\label{Comfo:EV}\\
    J^{cap,TH}_{u}=& \sum_{t \in \mathcal{T}} (\alpha^{cold}_u \Delta T^{cold}_{u,t} + \alpha^{hot}_u \Delta T^{hot}_{u,t}) && \forallu \label{Comfo:TH}
\end{align}
with $\Delta T^{cold} = max(0, min(T^{in,ref}\ut, T^{set}\ut)-T^{in}\ut)$ and $\Delta T^{hot} = max(0, T^{in}\ut-\bar{T}^{in,conf}_u)$.

\begin{align}  
&\Pinj\ut = \PPV\ut + \discharge\ut - \fix\ut - P^{flex}\ut - \charge\ut &\forallu, \forallt, \label{Exch:PbalPhys}\\ 
&\Pinj\ut = e^{ret}\ut + \exportCom\ut - i^{ret}\ut - \importCom\ut &\forallu, \forallt, \label{Exch:PbalVirt}\\
&\textstyle\sum_{u \in \mathcal{U}} \exportCom\ut = \sum_{u \in \mathcal{U}}\importCom\ut &\forallt. \label{Exch:ComBal}
\end{align}
with $\PPV\ut$ the photovoltaic power injection, $\discharge\ut$ and $\charge\ut$ the active powers going in and out of the battery storage system (BSS), $\fix\ut$ the non-flexible consumption and $\flex\ut$ the total flexible load consumption. Constraint (\ref{Exch:ComBal}) also ensures the export/import power balance within the REC.
}

\comment{
\begin{alignat}{2}\label{PB:model}
\!\min_{\Omega^U} \hspace{.5cm} & \text{CM Objective function \eqref{Obj:CM}} &\\
\text{subject to} \hspace{.5cm} & \textit{Cost equations} \hspace{.2cm} \eqref{Cost:CM} ,\notag\\
&\textit{Energy exchange constraints} \hspace{.2cm} \eqref{Exch:PbalPhys}-\eqref{Exch:ComBal}, \notag\\
&\textit{Flexibility activation constraints} \hspace{.2cm} \eqref{FA:Real}-\eqref{FA:Lim} \notag\\
&+ \eqref{FA:Eq1}-\eqref{FA:Eq3} \textit{ or } \eqref{FA:Pro1}-\eqref{FA:Pro2} \textit{ or } \eqref{FA:Cas1}-\eqref{FA:Cas3}, \notag\\
&\textit{Flexibility remuneration} \hspace{.2cm} \eqref{FR:Gain}-\eqref{FR:Distrib}, \notag\\
&\textit{Battery storage constraints} \hspace{.2cm} \eqref{BSS:Init}-\eqref{BSS:SOCFinal}, \notag\\
\Omega^L \in \;\;&\arg\min \text{REC Member Objective \eqref{Obj:Member}}\notag\\
\text{s.t.  }&\textit{Flexible devices constraints} \hspace{.2cm} \eqref{EV:Arr}-\eqref{HP:COP}, \notag\\
&\textit{Discomfort constraints}. \hspace{.2cm}
\eqref{Comfo:Tot}-\eqref{Comfo:TH}\notag
\end{alignat}

with \(\Omega^U:=\{i^{ret}\ut, \importCom\ut, e^{ret}\ut, \exportCom\ut, P^{\text{+,act}}\ut, P^{\text{-,act}}\ut, P^{cha}\ut,\) \( P^{dis}\ut, \pi^{up}_t, \pi^{down}_t\}\) the upper-level decision variables and

$\Omega^L:=\{P^{+,cap}\ut,P^{-,cap}\ut\}$ the lower-level decision variables.
\subsection{Objective functions}
CM Objective function:
\begin{equation}\label{Obj:CM}
    F(\Omega^U,\Omega^L) = \Pi^{REC}
\end{equation}
with $\Pi^{REC}$ the collective electricity bill in the community framework (\ref{Cost:CM}).

REC Member Objective function:
\begin{align}\label{Obj:Member}
    f_u(\Omega^U,\Omega^L) = &J^{cap,tot}_u \\
    & -\sum_{t\in\mathcal{T}}[\pi^{up}_t P^{+,cap}\ut+\pi^{down}_t P^{-,cap}\ut]\timestep \notag
\end{align}
with $J^{cap,tot}_u$ the total discomfort function of member $u$ (\ref{Comfo:Tot}) associated with the offered flexibility capacity and $\sum_{t\in\mathcal{T}}[\pi^{up}_t P^{+,cap}\ut+\pi^{down}_t P^{-,cap}\ut]\timestep$ the benefits associated with the flexibility capacity offered.

\subsection{Upper-Level Constraints}
\subsubsection{Cost structure}
The individual community bills and the collective bill are defined by:
\begin{align}  
\Pi^{REC}_u =  \sum_{t \in \mathcal{T}} &[\pi^{i,ret}i^{ret}\ut-\pi^{e,ret}e^{ret}\ut + \gamma^{com}(\importCom\ut+\exportCom\ut)]\timestep \label{Cost:Bill_u}
\end{align}
\begin{align}
\Pi^{REC} =  \sum_{u \in \mathcal{U}} \Pi^{REC}_u \label{Cost:CM}
\end{align}
where $i^{ret}\ut$ and $\importCom\ut$ (resp. $e^{ret}\ut$ and $\exportCom\ut$) represent the power imported from (resp. exported to) the retailer and the REC by user $u$ at time $t$. $\pi^{i,ret}$ and $\pi^{e,ret}$ are equivalent retailer import and export prices accounting for commodity and upstream grid usage charges. The commodity costs for the local energy exchanges are considered zero, the distribution of the benefits being done in the second step. However, a fee is charged for the intra-community exchanges, $\gamma^{com}$, accounting for the internal grid usage costs and the community operator remuneration. 
\subsubsection{Energy exchanges}
The power injected by member $u$ at time $t$ can be expressed as a function of physical (\ref{Exch:PbalPhys}) or virtual (\ref{Exch:PbalVirt}) flows:


\subsubsection{Flexibility Activation}\label{sec:flex_activation}
\begin{flalign}  
& \Pow{flex} = \Pow{ref,flex} + \Pow{+,act} - \Pow{-,act} \label{FA:Real}\\
&\Pow{ref,flex} = \Pow{ref,flex} + \Pow{ref,flex} \label{FA:Ref}\\
&\Pow{+/-,act} = \Pow{+/-,act} + \Pow{+/-,act}\label{FA:Act}\\
& \textbf{0} \leq \Pow{+/-,act} \leq  \Pow{+/-,cap} &\label{FA:Lim}
\end{flalign}
with $P^{flex}\ut$ the real consumption of flexible devices after flexibility activation, $P^{ref,flex}\ut$ the reference consumption of flexible devices, $P^{+,act}\ut$ and $P^{-,act}\ut$ the activated upward and downward flexibility either coming from electric vehicles ($\mathcal{E}_u$) or heat pumps ($\mathcal{HP}_u$). Constraint (\ref{FA:Lim}) limits the activated flexibility of devices in both directions to the offered flexible capacity $P^{+/-,cap}_{e_u/hp_u,t}$.

The flexibility activation is subject to one of the three following activation mechanism to ensure a fair contribution to each member:
\paragraph{Equal Activation}\hfill 
\begin{flalign}
&P^{+/-,act}\ut z^{+/-,cap}_{v,t} = P^{+/-,act}_{v,t} z^{+/-,cap}_{u,t} \hfill\\
&&\mathllap{\hfill \forall (u,v) \in \mathcal{U}\times\mathcal{U}, \forallt }\notag\\
&P^{+/-,cap}\ut \leq M\cdot z^{+/-,cap}\ut & \forall u, \forall t \\
&z^{+/-,cap}_{u,t} \in \{0,1\} & \forall u, \forall t
\end{flalign}

\subsection{Updated ISGT model}

\paragraph{Prorate Activation}

\begin{flalign}
&P^{+/-,act}\ut = P^{+/-,cap}\ut \textstyle\sum_{j \in \mathcal{J}} [0.01\cdot y^{+/-,cap}_{j,t}\cdot 2^{j-1}]&\label{FA:Pro1}\\
&&\mathllap{\forallu, \forallt}\notag\\
&y^{+/-,cap}_{j,t} \in \{0,1\} 
& \mathllap{\forall j \in \mathcal{J}, \forallt}\label{FA:Pro2}  
\end{flalign}
\paragraph{Cascade Activation}
\begin{align}
    &P^{+/-,cap}\ut z^{+/-,cap}\ut \leq P^{+/-,act}\ut& \forallu, \forallt\label{FA:Cas1}\\
    &P^{+/-,act}\ut \leq P^{+/-,act}_{v,t}+M\cdot z^{+/-,cap}_{v,t} & \forall (u,v) \in \mathcal{U}, \forallt\label{FA:Cas2}\\
    &z^{+/-,cap}_{u,t} \in \{0,1\} & \forallu, \forallt \label{FA:Cas3}
\end{align}

Furthermore, the real consumption of flexible devices after flexibility activation must satisfy the flexible devices constraints \eqref{}.

\subsubsection{Flexibility Remuneration}
\begin{flalign}  
& B^{REC} = \Pi^{REC,fixed} - \Pi^{REC} \label{FR:Gain}\\
& B^{REC} = \sum_{t\in\mathcal{T}}[\pi^{up}_t P^{+,act}\ut+\pi^{down}_t P^{-,act}\ut]\timestep&\label{FR:Distrib}
\end{flalign}
with $\Pi^{REC,fixed}$ the community bill without activation of flexibility, \textit{i.e.}, $P^{flex}\ut=P^{ref,flex}\ut$ and $B^{REC}$ the bill savings due to flexibility activation.

\subsection{Lower-Level Constraints}
\subsubsection{Flexible devices} REC members may consider shifting the consumption of certain appliances. This represents their flexibility potential. This work considers two types of flexible devices: electric vehicle chargers and heat pumps.

\textbf{Electric vehicles} are modeled as one-way batteries for which we can adjust the charging power. The charging operation is characterized by the arrival and departure times to or from the charging station.

\textbf{Heat pumps} 
are considered to be power-flexible loads which are used to meet a specific heating demand of buildings. The thermal behavior of buildings is modeled as a storage system as follows
\subsubsection{Flexibility capacity provision}
The offered flexibility of each device type can be expressed as:
\begin{align}  
&  P^{EV}_{e_u,t} = P^{ref,flex}_{e_u,t} + P^{+,cap}_{e_u,t} - P^{-,cap}_{e_u,t} \notag\\ 
&&\mathllap{\forall e_u \in \mathcal{E}_u, \forallt}\label{FC:EV}\\
&  P^{HP}_{hp_u,t} = P^{ref,flex}_{hp_u,t} + P^{+,cap}_{hp_u,t} - P^{-,cap}_{hp_u,t} \notag\\ 
&& \mathllap{\forall hp_u \in \mathcal{HP}_u, \forallt}\label{FC:HP}\\
&  0 \leq P^{+,cap}_{e_u/hp_u,t} \leq \bar{P}^{EV/HP}_{e_u/hp_u}-P^{ref,flex}_{e_u/hp_u,t} \notag\\ 
&& \mathllap{\forall e_u/hp_u \in \mathcal{E}_u/\mathcal{HP}_u, \forallt}\label{FC:Up_max}\\
&  0 \leq P^{-,cap}_{e_u/hp_u,t} \leq P^{ref,flex}_{e_u/hp_u,t} \notag\\ 
&& \mathllap{\forall e_u/hp_u \in \mathcal{E}_u/\mathcal{HP}_u, \forallt}&&&&&&&&&&&\label{FC:Up_min}
\end{align}
The overall upward $P^{+,cap}_{u,t}$ and downward $P^{-,cap}_{u,t}$ capacity offered by a member are the sum of both contribution from electric vehicles and heat pumps.
}

\bibliographystyle{IEEEtran}
\bibliography{references}
\end{document}

%% file: macros.tex

\newcommand{\BSSs}{\ensuremath{\mathcal{D}^{sto}}}
\newcommand{\OPSOC}{\ensuremath{s}}
\newcommand{\maxcharge}{\ensuremath{\overline{s}}}
\newcommand{\mincharge}{\ensuremath{\underline{s}}}
\newcommand{\chargerate}{\ensuremath{\overline{P}^{\text{cha}}}}
\newcommand{\dischargerate}{\ensuremath{\overline{P}^{\text{dis}}}}
\newcommand{\retentionRate}{\ensuremath{\eta^{\text{retention}}}}
\newcommand{\chargeEfficiency}{\ensuremath{\eta^{\text{cha}}}}
\newcommand{\dischargeEfficiency}{\ensuremath{\eta^{\text{dis}}}}
\newcommand{\initialCharge}{\ensuremath{s^\text{init}}}
\newcommand{\minEndCharge}{\ensuremath{\underline{S}^\text{end}}}
\newcommand{\maxEndCharge}{\ensuremath{\overline{S}^\text{end}}}
\newcommand{\charge}{\ensuremath{\text{P}^\text{cha}}}
\newcommand{\discharge}{\ensuremath{\text{P}^\text{dis}}}
\newcommand{\finalCharge}{\ensuremath{S^{\text{end}}}}
\newcommand{\BSSFee}{\ensuremath{\gamma^\text{sto}}}

\newcommand{\devices}[1]{\ensuremath{\mathcal{D}^\text{#1}}}
\newcommand{\sheddableDevices}{\ensuremath{\devices{she}}}
\newcommand{\nonflexibleDevices}{\ensuremath{\devices{nfl}}}
\newcommand{\steerableDevices}{\ensuremath{\devices{ste}}}
\newcommand{\curtableDevices}{\ensuremath{\devices{nst}}}
\newcommand{\nonsteerableDevices}{\ensuremath{\devices{nst}}}

\newcommand{\users}{\ensuremath{\mathcal{U}}}
\newcommand{\nodes}{\ensuremath{\mathcal{N}}}

\newcommand{\bill}[1]{\ensuremath{\text{B}^\text{#1}}}
\newcommand{\optprofit}[1]{\ensuremath{B^{*, \text{#1}}}}

\newcommand{\profit}[1]{J^\text{#1}}
\newcommand{\profitshare}[2]{\ensuremath{r^{#1}(J^{*, \text{MU}}, #2)}}

\newcommand{\maxExportToGrid}{\ensuremath{E}_{u,t}^{\text{cap}}}
\newcommand{\maxImportFromGrid}{\ensuremath{I}_{u,t}^{\text{cap}}}

\newcommand{\operatorincome}{\ensuremath{J}^\text{operator}}
\newcommand{\OPFee}{\ensuremath{\gamma^\text{com}}}

\newcommand{\peak}{\ensuremath{\overline{p}}}

\newcommand{\price}[1]{\ensuremath{\pi^{\text{#1}}}}
\newcommand{\sheddingPrice}{\price{she}}
\newcommand{\flexiblePrice}{\price{flex}}
\newcommand{\reservePrice}{\price{res}}
\newcommand{\gridBuyPrice}{\price{i,ret}}
\newcommand{\gridSalePrice}{\price{e,ret}}
\newcommand{\lossPrice}{\price{loss}}

\newcommand{\OPcost}[1]{\ensuremath{c^\text{#1}}}

\newcommand{\exportGrid}{\ensuremath{\text{e}^\text{ret}}}
\newcommand{\importGrid}{\ensuremath{\text{i}^\text{ret}}}
\newcommand{\exportCom}{\ensuremath{\text{e}^\text{com}}}
\newcommand{\importCom}{\ensuremath{\text{i}^\text{com}}}
\newcommand{\Pprod}{\ensuremath{\text{p}^{\text{prod}}}}
\newcommand{\Pcons}{\ensuremath{\text{p}^{\text{cons}}}}

\newcommand{\OPreserve}[1]{\ensuremath{r^\text{#1}}}
\newcommand{\reserveBSSInc}{\ensuremath{r^{s+}_{d,t}}}
\newcommand{\reserveBSSDec}{\ensuremath{r^{s-}_{d,t}}}

\newcommand{\timestep}{\ensuremath{\Delta_T}}
\newcommand{\OPperiods}{\ensuremath{\mathcal{T}}}
\newcommand{\OPdays}{\ensuremath{\mathcal{D}}}

\newcommand{\flexRamp}{\ensuremath{\Delta \overline{P}^\text{flex}}}
\newcommand{\fix}{\ensuremath{\text{P}^\text{fix}}}
\newcommand{\PPV}{\ensuremath{\text{P}^\text{PV}}}
\newcommand{\flexible}{\ensuremath{\overline{P}^\text{flex}}}
\newcommand{\PPVprof}{\ensuremath{\overline{\text{P}}^\text{PV}}}
\newcommand{\PVprof}{\ensuremath{\overline{p}^\text{PV}}}


\newcommand{\flex}{\ensuremath{\text{P}^\text{flex}}}
\newcommand{\Pts}{\ensuremath{\text{P}^\text{TS}}}
\newcommand{\Pev}{\ensuremath{\text{P}^\text{EV}}}
\newcommand{\Pth}{\ensuremath{\text{P}^\text{TH}}}

\newcommand{\flexcoef}{k^{\text{flex}}}
\newcommand{\disccoef}{k^{\text{disc}}}

\newcommand{\discomf}[1]{\ensuremath{\text{J}^\text{#1}}}

\newcommand{\Ecap}{\ensuremath{E^\text{cap,flex}}}
\newcommand{\Eact}{\ensuremath{E^\text{act,flex}}}
\newcommand{\actPrice}{\ensuremath{\pi^\text{act}}}
\newcommand{\capPrice}{\ensuremath{\pi^\text{cap}}}
\newcommand{\flexPrice}{\ensuremath{\pi^\text{flex}}}

\newcommand{\shed} {\ensuremath{a^\text{she}}}
\newcommand{\flexCons}{\ensuremath{c^\text{flex}}}
\newcommand{\flexStart}{\ensuremath{y^\text{flex}}}
\newcommand{\flexStarted}{\ensuremath{z^\text{flex}}}
\newcommand{\exclusiveGroup}{\ensuremath{\mathcal{G}}}
\newcommand{\nonFlexible}{\ensuremath{D^\text{nfl}}}
\newcommand{\flexibleLoadDuration}{\ensuremath{\text{duration}^\text{fl}}}
\newcommand{\flexibleLoadProfile}{\ensuremath{\text{profile}^\text{flex}}}
\newcommand{\flexibleLoadStartTime}{\ensuremath{\text{start}^\text{flex}}}
\newcommand{\flexibleLoadEndTime}{\ensuremath{\text{end}^\text{flex}}}
\newcommand{\flexibleLoadAcceptanceRatio}{\ensuremath{\underline{a}^\text{flex}}}
\newcommand{\flexibleLoadMustRun}{\ensuremath{\text{run}^\text{flex}}}
\newcommand{\flexibleEnergy}{\ensuremath{\text{E}^\text{flex}}}
\newcommand{\sheddable}{\ensuremath{D^\text{she}}}
\newcommand{\maxSheddingTime}{\ensuremath{\overline{A}^\text{she}}}

\newcommand{\InvHor} {\ensuremath{I^{\text{hor}}}}

\newcommand{\CPV} {\ensuremath{C^\text{PV}}}
\newcommand{\CBSS} {\ensuremath{C^\text{BSS}}}

\newcommand{\hasPV} {\ensuremath{b^\text{PV}}}
\newcommand{\hasBSS} {\ensuremath{b^\text{BSS}}}
\newcommand{\hasdiscount}[1] {\ensuremath{d^\text{#1}}}

\newcommand{\pif}[1] {\ensuremath{\pi_f^\text{#1}}}
\newcommand{\piv}[1] {\ensuremath{\pi_v^\text{#1}}}
\newcommand{\pid}[1] {\ensuremath{\pi_d^\text{#1}}}

\newcommand{\Cmax}[1] {\ensuremath{\overline{C}^\text{#1}}}
\newcommand{\Cmin}[1] {\ensuremath{\underline{C}^\text{#1}}}

\newcommand{\weight}[1]{\ensuremath{w^\text{#1}}}

\newcommand{\foralld}{\ensuremath{\forall d \in \OPdays}}
\newcommand{\forallt}{\ensuremath{\forall t \in \OPperiods}}
\newcommand{\foralln}{\ensuremath{\forall n \in \nodes}}
\newcommand{\forallu}{\ensuremath{\forall u \in \users}}
\newcommand{\forallut}{\ensuremath{\forall u \in \users, t \in \OPperiods}}
\newcommand{\foralluPrime}{\ensuremath{\forall u' \in \users}}
\newcommand{\forallb}{\ensuremath{\forall d \in \BSSs}}
\newcommand{\forallDShed}{\ensuremath{\forall d \in \sheddableDevices}}
\newcommand{\forallDFlex}{\ensuremath{\forall d \in \devices{fl}}}
\newcommand{\forallDCurt}{\ensuremath{\forall d \in \devices{nst}}}
\newcommand{\forallDSteer}{\ensuremath{\forall d \in \devices{ste}}}

\newcommand{\uinU}{u \in \users}
\newcommand{\tinT}{t \in \OPperiods}

\newcommand{\ut}{\ensuremath{_{u,t}}}
\newcommand{\ud}{\ensuremath{_{u,d}}}

\newcommand{\Rplus}{\ensuremath{\mathbb{R}_+}}
\newcommand{\Rminus}{\ensuremath{\mathbb{R}_-}}

\newcommand{\Ploss}{\ensuremath{P^\text{loss}}}
\newcommand{\Pinj}{\ensuremath{\text{P}^\text{inj}}}
\newcommand{\Qinj}{\ensuremath{Q^\text{inj}}}
\newcommand{\Qi}[1]{\ensuremath{Q^\text{#1}}}
\newcommand{\Isqr}{\ensuremath{I^\text{sqr}}}
\newcommand{\Vsqr}{\ensuremath{V^\text{sqr}}}

\newcommand{\Pline}{\ensuremath{P^\text{line}}}
\newcommand{\Qline}{\ensuremath{Q^\text{line}}}

\newcommand{\Vmax}{\ensuremath{\overline{V}}}
\newcommand{\Vmin}{\ensuremath{\underline{V}}}
\newcommand{\Imax}{\ensuremath{\overline{I}}}

\newcommand{\lines}{\ensuremath{\mathcal{L}}}
\newcommand{\child}{\ensuremath{\mathcal{C}}}
\newcommand{\ancestor}{\ensuremath{\mathcal{A}}}

\newcommand{\red}[1]{\textcolor{red}{#1}}
\newcommand{\todo}[1]{\textcolor{red}{\textbf{TO DO: #1}}}
\newcommand{\NB}[1]{\textcolor{blue}{NB: #1}}
\renewcommand{\comment}[1]{}

\newcommand{\UxT}{\ensuremath{\mathcal{U}\times\mathcal{T}}}
\newcommand{\SoC}[1]{\ensuremath{\text{s}^{\text{#1}}}}
\newcommand{\SoCmax}[1]{\ensuremath{\bar{\text{s}}^{\text{#1}}}}
\newcommand{\SoCmin}[1]{\ensuremath{\text{\underline{s}}^{\text{#1}}}}
\newcommand{\Capa}[1]{\ensuremath{\text{C}^{\text{#1}}}}
\newcommand{\eff}[1]{\ensuremath{\eta^{\text{#1}}}}
\newcommand{\Pmax}[1]{\ensuremath{\bar{\text{P}}^{\text{#1}}}}
\newcommand{\Bool}{\ensuremath{\{0,1\}}}
\newcommand{\Temp}[1]{\ensuremath{\text{T}^{\text{#1}}}}
\newcommand{\Tempmax}[1]{\ensuremath{\bar{\text{T}}^{\text{#1}}}}
\newcommand{\Pow}[1]{\ensuremath{\text{P}^{\text{#1}}}}
\newcommand{\Powprim}[1]{\ensuremath{\text{P}'^{\text{#1}}}}
\newcommand{\degC}{\ensuremath{^\circ\mathrm{C}}}
\newcommand{\diag}[1]{\ensuremath{\operatorname{diag}(#1)}}